# The Effects of Temperature Acclimation on Swimming Performance in the Pelagic Mahi-Mahi (*Coryphaena hippurus*)

Rachael M. Heuer[1]\*, John D. Stieglitz[1], Christina Pasparakis[1†], Ian C. Enochs[2], Daniel D. Benetti[1] and Martin Grosell[1]

[1] Department of Marine Biology and Ecology, University of Miami, Rosenstiel School of Marine and Atmospheric Sciences, Miami, FL, United States, [2] NOAA, Atlantic Oceanographic and Meteorological Laboratory, Ocean Chemistry and Ecosystem Division, Miami, FL, United States

Mahi-mahi (*Coryphaena hippurus*) are a highly migratory pelagic fish, but little is known about what environmental factors drive their broad distribution. This study examined how temperature influences aerobic scope and swimming performance in mahi. Mahi were acclimated to four temperatures spanning their natural range (20, 24, 28, and 32°C; 5–27 days) and critical swimming speed ($U_{crit}$), metabolic rates, aerobic scope, and optimal swim speed were measured. Aerobic scope and $U_{crit}$ were highest in 28°C-acclimated fish. 20°C-acclimated mahi experienced significantly decreased aerobic scope and $U_{crit}$ relative to 28°C-acclimated fish (57 and 28% declines, respectively). 32°C-acclimated mahi experienced increased mortality and a significant 23% decline in $U_{crit}$, and a trend for a 26% decline in factorial aerobic scope relative to 28°C-acclimated fish. Absolute aerobic scope showed a similar pattern to factorial aerobic scope. Our results are generally in agreement with previously observed distribution patterns for wild fish. Although thermal performance can vary across life stages, the highest tested swim performance and aerobic scope found in the present study (28°C), aligns with recently observed habitat utilization patterns for wild mahi and could be relevant for climate change predictions.

**Keywords: respirometry, climate change, $U_{crit}$, dolphinfish, swim tunnel, metabolic rate, thermal, aerobic scope**

## INTRODUCTION

Temperature is a vital factor dictating swimming performance and aerobic capacity in fish (Clark et al., 2013). In one conceptual model, the oxygen and capacity limited thermal tolerance (OLCTT), an animal performs optimally across a range of temperatures where aerobic scope is maximized (between upper and lower pejus temperatures). As temperatures increase or decrease from pejus temperatures, a growing discrepancy between oxygen supply and demand ultimately causes a decline in organismal performance (Pörtner and Knust, 2007; Pörtner and Farrell, 2008;

Pörtner, 2010). Recent studies have also suggested that oxygen-independent mechanisms such as protein dysfunction, or muscle function may be responsible for declines in performance (Ern et al., 2014, 2016; Slesinger et al., 2019). Support for oxygen-independent limitations in performance has also been supported in recent work examining oxygen supply capacity (Seibel and Deutsch, 2020). Regardless of the mechanism, physiological performance is constrained during exposure to extreme temperature scenarios. Thus, fish with broad geographic distributions must be able to maintain performance over a wide range of temperatures or possess the ability to select optimal thermal environments within their distribution range.

Mahi-mahi (*Coryphaena hippurus*) (referred to herein as "mahi") are an ecologically and economically important species that are circum-globally distributed (Palko et al., 1982). Mahi have large metabolic demands and are highly migratory, grow rapidly, and spawn frequently (Palko et al., 1982; Benetti et al., 1995a; Brill, 1996; Oxenford, 1999; Merten et al., 2014a; Stieglitz et al., 2017; Schlenker et al., 2021). Most life stages can be found in either open pelagic waters or associated with Sargassum brown algal mats floating in the surface (Casazza and Ross, 2008). Sargassum mats are thought to play an important role for not only early life stages, but also serve as hunting grounds for juveniles and adults (Casazza and Ross, 2008; Kitchens and Rooker, 2014). Adult mahi spend most of their time near the surface of the ocean but will make vertical excursions deeper into the water column (Furukawa et al., 2011). Few studies have specifically aimed to quantify the biology and movement of fish in the transition from the juvenile to early adult stage. However, mahi in this size range (10–40 cm standard length) have been found in the Gulf Stream and the Florida Current (Oxenford, 1999), and begin to show adult morphology, coloration, and schooling behavior as early as ∼45 days post hatch (Perrichon et al., 2019).

Most of our knowledge regarding the thermal niche in which mahi reside originates from fisheries reporting, ichthyoplankton surveys, and data collected using pop-up satellite archival tags (PSATs). Mahi do not consistently reside in waters below 20°C (Flores et al., 2008; Martínez-Rincón et al., 2009; Furukawa et al., 2011, 2014; Farrell et al., 2014; Merten et al., 2014b; Lin et al., 2019; Schlenker et al., 2021), and the upper end of their range is 30–31°C (Hammond, 2008; Farrell et al., 2014; Merten et al., 2014b; Schlenker et al., 2021), although one study reports larval mahi in waters up to ∼33°C (Kitchens and Rooker, 2014). Exposure to this large range of temperatures is not a surprise considering mahi perform extensive migrations and make frequent vertical excursions in the water column (Merten et al., 2014b, 2016; Schlenker et al., 2021). Despite access to a broad thermal range, average temperatures where mahi reside are more constrained. Mahi catch rates are typically higher in warmer sea surface temperatures ranging from 24 to 29°C (Flores et al., 2008; Martínez-Rincón et al., 2009; Farrell et al., 2014). This reported range is supported by mahi tagging data, which indicates this species occupies waters with mean temperatures of 25.05–28.9°C in the western Atlantic (Hammond, 2008; Merten et al., 2014b). In one PSAT study, the median preferred temperature of 19 fish tagged throughout the Gulf of Mexico and the western Atlantic was 27.5°C and 95% of their time was spent between 25 and 29°C (Schlenker et al., 2021). Larval mahi have been found year round in the Gulf of Mexico using ichthyoplankton surveys, but tend to be caught more frequently at temperatures above 24°C but below 29°C (Ditty et al., 1994; Kitchens and Rooker, 2014). Also, predicted spawning events for wild fish primarily occurred between 27.5 and 30°C, which provides some insight in the habitat into which early development occurs (Schlenker et al., 2021).

While our understanding of mahi thermal habitat utilization continues to expand, our understanding of what physiological constraints may underlie their distribution remains limited. Swim tunnel respirometry has long been used as a tool to assess the relationship between fitness and temperature (Keen and Farrell, 1994; Claireaux et al., 2006; Clark et al., 2013). In one swim tunnel study on mahi, acute temperature elevation from 27 to 30°C, (1°C/h) led to increased mortality in juveniles, but surviving individuals showed no change in swimming performance (Mager et al., 2018). In another study examining responses to rapid warming in early juvenile mahi (0.83–1.09°C/h), bottom-sitting and mortality were observed at temperatures ranging from 31.7 to 33°C. Although heating was not extended beyond 33°C, these findings suggest that mahi at this life stage experience significant challenges beyond this temperature (Szyper and Lutnesky, 1991).

In an effort to quantify performance capacity of mahi across a range of temperatures, we measured standard and maximum metabolic rate, aerobic scope, critical swimming speed ($U_{crit}$), optimal swimming speed ($U_{opt}$), and cost of transport (COT) in mahi acclimated to 20, 24, 28, or 32°C, covering their natural range. Since mahi are known to show adult coloration and schooling behavior as early as 45 days post hatch, the life stage utilized in the present study (mean of 61–70 days post hatch) represented late juveniles and/or juveniles transiting to an early adult life stage. Since all life stages are typically and most often found in surface waters, and at the same geographical distribution, we cautiously compare findings from this study to recent patterns noted in wild larger adult mahi fitted with PSATs (Schlenker et al., 2021) that are logistically difficult to obtain and test using swim tunnel respirometry.

In addition to these lab-based measurements potentially providing insight into habitat utilization patterns, this study could also help us understand how mahi may respond to global projected temperature increases for the present century of 1–4°C (Masson-Delmotte et al., 2018). In the Gulf of Mexico, current mean summer sea surface temperatures are ∼29.5°C, but are projected to reach >32°C by year 2100 (Laurent et al., 2018). Such changes could introduce a selective pressure, and/or lead to alterations in habitat use if swim performance and aerobic capacity decline at high temperatures. Limited laboratory data (Szyper and Lutnesky, 1991; Mager et al., 2018), and PSAT data showing that mahi spend little time beyond 31°C despite access to warmer temperatures (Schlenker et al., 2021) suggests these changes may already be occurring. Given current knowledge from catch rates and PSATs, we hypothesized that mahi acclimated to the lowest (20°C) and the highest (32°C) temperatures would show declines in aerobic scope and swimming performance.

## MATERIALS AND METHODS

### Experimental Animals

Wild mahi-mahi (*C. hippurus*) were caught offshore of Miami, FL, United States and transferred to land-based maturation and spawning tanks (15,000-L volume) at the University of Miami Experimental Hatchery (UMEH) for use as broodstock (Stieglitz et al., 2017). F1 generation offspring used in this study were reared from egg to the late juvenile/early adult size using UMEH aquaculture techniques involving the feeding of live planktonic prey during early stages (enriched rotifers and *Artemia nauplii*) with eventual transition to a pelletized diet (Otohime, Reed Mariculture). In a recent publication, mahi were considered to be at the juvenile stage for a long period of development ∼16–55 days post hatch. However, by 45 days post hatch, mahi show typical adult coloration and exhibit schooling behavior, and by 55 days post hatch, the tail is fully forked and they are considered to be in a "transition to young adult phase." Thus, fish used in the present study (mean 61–70 days post hatch) were considered to be late juveniles or early adults. In optimal rearing conditions, sexual maturity is reached at 80–90 days post hatch (Perrichon et al., 2019). Fish were maintained in flow-through seawater (22–27°C) throughout the rearing process. This range of rearing temperatures is similar to the range of temperatures at which early life stage mahi are found in the wild (Ditty et al., 1994; Kitchens and Rooker, 2014). All animal care and experimental protocols were reviewed, approved, and carried out in accordance with relevant guidelines for experiments on teleost vertebrates provided by the University of Miami's Institutional Animal Care and Use Committee (IACUC protocol 15-019).

### Experimental Tanks and Temperature Acclimation

Tanks for temperature acclimation were used as described in Enochs et al. (2018). Each tank replicate was comprised of one 75 L acrylic top aquaria and a connected 75 L sump tank on the bottom. Seawater pumped from the bottom to the top ensured constant circulation and facilitated flow in the top tank. Seawater from Bear Cut, Miami, FL continuously flowed into each tank replicate system (∼500 mL/min) and was aerated in the bottom tank. The incoming flow of fresh seawater was controlled by needle valves and monitored by optical gate flow meters (Micro-Flo, Serv-A-Pure). Temperature in the tanks was controlled using a digital feedback system where temperature was monitored in the top tank using a high-accuracy RTD sensor (TTD25C, ProSense) and heating/chilling occurred in the bottom sump tank. Heating was achieved using a 300 watt aquarium heater (TH-300, Finnex) or an 800 watt heater for the 32°C temperature treatment (TH-800, Finnex). Cooling was achieved using a titanium chiller coil (Hotspot Energy) attached to a sealed cold water source with an electronically actuated solenoid valve. Both seawater flow and temperature were controlled at a central computer via a USB connection through a modular hardware interface (CompactDAQ, National Instruments) and using software written in LabVIEW (National Instruments) (Enochs et al., 2018). Mahi in the tanks were placed on a 10:14 light-dark cycle and tank lighting was controlled using 135W LED arrays (Hydra 52 HD, Aqua Illumination) located directly above the top tank.

Juvenile mahi were transferred from the UMEH and randomly assigned to one of four temperature treatments, 20, 24, 28, or 32°C ($n$ = 3–6 tank replicates per temperature, 8–13 fish per tank). Mahi were first placed in experimental tanks set at the temperature of the rearing tanks (∼22–25°C) and the experimental tanks were gradually ramped over 48 h to their respective target temperatures. Mahi were exposed to temperature treatments for 5–27 days, with a mean exposure period of 12.9 ± 1.6, 12.2 ± 1.4, 10.8 ± 1.3, and 9.0 ± 1.5 days, for 20, 24, 28, and 32°C (Mean ± SEM), respectively. Mahi were fed a pelletized diet daily *ad libitum* (Otohime, Reed Mariculture), however, food was withheld 24–36 h prior to commencement of swim trials to eliminate the impact of digestive metabolism on swim chamber respirometry studies (Stieglitz et al., 2018). Mass, fork length, and age of fish used for each temperature treatment are presented in **Table 1**.

Water quality parameters were measured daily in experimental tanks (**Supplementary Table 1**) and included salinity, temperature, dissolved oxygen, and $pH_{NBS}$. Salinity and temperature were recorded using a WTW 3310 meter connected to a TetraCon 325 probe. $pH_{NBS}$ was measured using a Hach H160 meters attached to a Radiometer PHC3005 electrode. Dissolved oxygen was recorded using a YSI meter and ProODO optical probe. Water samples were collected three times per week for total ammonia and measured using a micro-modified colorimetric assay (Ivancic and Degobbis, 1984).

### Swimming Performance

Five-liter Brett-style swim tunnel respirometers (Loligo Systems, Denmark) were used to assess oxygen consumption and critical swimming speed ($U_{crit}$) in mahi (Stieglitz et al., 2016, 2018) at the four treatment temperatures mentioned above. Automated control of intermittent respirometry (20-min measurement loops: divided into flush, stabilization, and measurement periods) was achieved using AutoResp 2.1.0 software (Loligo Systems, Denmark). Typically, during these 20 min (1,200 s) loops, the flush cycle was 870 s, the wait cycle was 30 s, and the measurement period was 300 s. While the wait cycle always remained at 30 s, the measurement and flush cycles would sometimes be adjusted to have maximum resolution (Rosewarne et al., 2016), while ensuring fish never experienced air saturation below 80%. For example, during the fastest portion of the $U_{crit}$ test or

**TABLE 1 |** Biometric data for mahi-mahi used in swim tunnel experiments.

| Treatment temperature (°C) | n | Body mass (g) | Fork length (cm) | Age (days) |
|---|---|---|---|---|
| 20 | 16 | 32.8 ± 1.2 | 15.1 ± 0.2 | 66.4 ± 2.0 |
| 24 | 14 | 37.1 ± 2.9 | 15.3 ± 0.4 | 61.1 ± 2.6 |
| 28 | 14 | 39.8 ± 2.5 | 15.6 ± 0.4 | 61.0 ± 2.8 |
| 32 | 13 | 39.6 ± 4.7 | 15.2 ± 0.7 | 70.0 ± 4.6 |

*Values represent means ± SEM.*

during introduction of fish into the swim tunnel, the flush cycle would be 970 s, and the measurement cycle would be 200 s. The rate of oxygen depletion during the measurement cycle was automatically calculated in real-time using AutoResp 2.1.0 software, and air saturation and $r^2$ values were carefully monitored. Oxygen was measured using a Pt100 fiber-optic mini-sensor probe connected to a Witrox 1 oxygen meter (PreSens Precision Sensing, Germany). Calibration of oxygen sensors was performed prior to introducing mahi to the swim tunnels as in Stieglitz et al. (2016). Briefly, a two-point calibration was achieved by using 100% air-saturated UV-sterilized seawater as the high calibration point and a solution of 10 g L$^{-1}$ Na$_2$SO$_3$ as the low (zero saturation) calibration point. Water velocity was calibrated using a handheld anemometer coupled with a 30-mm cylinder probe vane wheel flow sensor (Höntzch GmbH) to measure speed (cm s$^{-1}$) at different motor output settings (Hz). Temperatures in the swim tunnels were matched to respective treatment conditions (20, 24, 28, or 32°C) using a heater (Finnex Digital Heater Controller HC-810M with 800W titanium heater). Swim tunnels were on continuous flow-through during the swim tunnel acclimation and swim experiments, using aerated, 1-μm filtered seawater.

Prior to the commencement of swim trials, mahi were introduced to swim tunnels at a speed of one body length per second (BL s$^{-1}$). This acclimation speed was chosen based on previous studies since mahi are ram ventilators and would not survive in respirometers without significant flow (Stieglitz et al., 2016, 2018). An overnight acclimation period was found to be sufficient in obtaining stable oxygen consumption rates reflective of routine metabolic rate (RMR). Swim trials were conducted the following morning. $U_{crit}$ was performed by increasing water velocity from the 1.0 body lengths s$^{-1}$ acclimation speed in increments of 0.5 body lengths s$^{-1}$ every 20 min until the fish was fatigued. Fatigue was defined as continuous brushing with the tail at the rear screen, sustained resting on the caudal fin at the rear screen, or by the fish becoming pinned sideways against the rear screen of the tunnel (Mager et al., 2014; Stieglitz et al., 2016). Fatigue was determined in real-time using video relay to prevent disturbance during the test. $U_{crit}$, expressed in body length s$^{-1}$, was calculated using the Equation (1) determined by Brett (1964): $U_{crit} = [U_f + (T/t)dU]/cm$, where $U_f$ (cm s$^{-1}$) is the highest swim velocity maintained for a complete step interval, $T$ (s) is the length of time spent at the final, highest swim velocity, $t$ (s) is the set time for the step interval, $dU$ (cm s$^{-1}$) is the increment in swim speed of each step, and cm is the fork length of the fish. During swimming tests, a solid blocking correction was applied to account for differences in water velocity due to the size of the animal using AutoResp software (Loligo Systems, Denmark) (Bell and Terhune, 1970). The number of animals that were disqualified from the swimming test or died shortly after introduction to the swim tunnel was quantified (**Supplementary Figure 1**). The criteria for disqualification was either an $r^2$ of less than 0.70 as in previous studies on mahi (Mager et al., 2014; Stieglitz et al., 2016) in plots described in the next section and/or an animal that refused to engage in typical swimming behavior. This occurred in all treatments except for the 28°C-acclimated mahi. Mortality following introduction to the tunnel only occurred in the 32°C-acclimated animals ($n$ = 4 of 21 attempts; **Supplementary Figure 1**). Final sample sizes for determining $U_{crit}$ and metabolic rates are summarized in **Table 1**. Sample traces of respirometry trials can be found in **Supplementary Figure 2**, and oxygen partial pressure (kPa) during the final full swim interval can be found in **Supplementary Figure 3**.

## Metabolic Rates and Cost of Transport

The oxygen consumption rate was derived from the slope of the linear regression of oxygen content over time and automatically calculated using AutoResp software (Loligo Systems, Denmark). Aerobic scope is defined as the difference between maximum and standard metabolic rate. To determine these parameters, the logarithm of oxygen consumption (mg O$_2$ kg$^{-1}$ h$^{-1}$) was plotted as a function of swimming speed (body length s$^{-1}$) and a least squares linear regression was performed for each fish (Mager et al., 2014; Stieglitz et al., 2016). A sample trace is illustrated in **Supplementary Figure 4**. Standard metabolic rate (SMR) was the y-intercept and maximum metabolic rate (MMR) was the metabolic rate at $U_{crit}$. Absolute aerobic scope was determined as MMR-SMR and only regressions with an $r^2 \geq 0.7$ were used in the analyses. Data were normalized for body mass by scaling SMR and MMR values predicted for a 40 g fish using scaling coefficients previously recorded for this species (aerobic scope = normalized MMR - normalized SMR, SMR = 0.6088, MMR = 0.7936) (Stieglitz et al., 2016). A 40 g fish was chosen because this was close to the average mass for all fish used in the present study (37.1 g). Scaling coefficients were necessary to use because mahi in the age range used in the present study are known to have very fast growth rates [specific growth rate of 10% of their body per day (Benetti et al., 1995b)]. Although body mass varied, means were similar across temperature treatments. In addition to absolute aerobic scope, factorial aerobic scope was also reported (Clark et al., 2013) and was calculated by dividing MMR by SMR. COT was determined by dividing the oxygen consumption (MO$_2$) by swimming velocity. These values were plotted as a function of increasing swim velocity and fitted with a second order ($k$ = 2) polynomial regression model. This approach provided the minimum cost of transport (COT$_{min}$) and the cost of transport at $U_{crit}$ (COT$_{Ucrit}$). Further, optimal swimming speed ($U_{opt}$), which is the speed at which swimming a given distance required the minimum oxygen consumption was determined by fitting the first derivative of the polynomial to zero (Palstra et al., 2008; Stieglitz et al., 2016; see **Supplementary Figure 5** for representative trace). Only individuals with an $r^2$ of greater than or equal to 0.7 were used for analysis ($n$ = 12, 12, 12, 9, for 20, 24, 28, and 32°C, respectively). Mass was normalized to a 40 g fish using previously determined scaling coefficients in this species for COT$_{min}$ and COT$_{Ucrit}$ (COT$_{min}$ = 0.79, COT$_{Ucrit}$ = 0.59) (Stieglitz et al., 2016). Figures with without mass scaling coefficients can be found in **Supplementary Figure 6**.

Temperature coefficients (Q$_{10}$) were calculated at increasing temperatures for each tested endpoint using the equation, Q$_{10}$ = (k$_2$−k$_1$)$^{10/(t2-t1)}$, where k$_1$ and k$_2$ are the values recorded at the temperatures, $t_1$ and $t_2$, respectively (Kieffer et al., 1998). $Q_{10}$ values are presented in **Supplementary Table 2**.

## Statistical Analysis

Differences between temperature treatments were assessed using one-way ANOVAs in cases where data were normally distributed, followed by Holm-Sidak *post hoc* tests to perform pairwise comparisons between temperatures. Shapiro–Wilk tests were used to assess normality. Data that were not normally distributed were analyzed using Kruskal–Wallace ANOVA on ranks tests and *post hoc* pairwise comparisons were performed using Dunn's tests. A Welch's ANOVA was used when data violated the assumption of homogeneity of variance. Significance in the Welch ANOVA was followed by a Games-Howell *post hoc* test to compare between temperatures (McDonald, 2009). Significance for all statistical tests was determined at $p < 0.05$, and means are presented ± SEM. All statistics were performed in Sigmaplot 13.0, except for the Welch's ANOVA and associated *post hoc* tests, which were performed in SPSS (V26). Statistical comparisons and associated *p*-values are presented in **Supplementary Table 3**.

# RESULTS

## Mortality and Disqualifications

At the highest temperature (32°C), 20% of mahi died shortly after being transferred into the swim tunnel and another 20% were disqualified from testing, due to violation of the pre-established stipulations detailed in the methods section. Thus, only 60% of animals tested at this temperature successfully completed the test (**Supplementary Figure 1**). No mortality was noted in the other temperature treatments. Thirteen and 16% of mahi were disqualified from testing at 20 and 24°C, respectively, while no animals were disqualified at 28°C (**Supplementary Figure 1**).

## Metabolic Rate

Standard metabolic rate was significantly impacted by temperature (**Figure 1A**; $p < 0.001$). Although not statistically significant, the largest incremental increase occurred in the transition from 28 to 32°C, where standard metabolic rate increased by 44% ($Q_{10} = 2.49$, P = 0.219). Maximum metabolic rate was also significantly impacted by temperature ($p < 0.001$) and increased from 20 to 28°C, however, the largest incremental increase occurred between 20 and 24°C (**Figure 1B**, 52%, $Q_{10} = 2.86$, P = 0.004). There was no significant difference in maximum metabolic rate between 28 and 32°C ($p = 0.949$). Aerobic scope was significantly impacted by temperature (**Figure 1C**, $p < 0.001$). Aerobic scope was highest at 28°C and declined in either direction from this temperature and animals acclimated to 20°C had the lowest aerobic scope. This was significantly different from all other temperatures and 57% lower than the aerobic scope of 28°C-acclimated animals ($p < 0.001$). Driven largely by increased standard metabolic rate, mahi acclimated to 32°C experienced a non-significant 10% decline in aerobic scope compared to 28°C ($Q_{10} = 0.76$, $p = 0.328$). Factorial aerobic scope showed a similar pattern to aerobic scope and was influenced significantly by temperature, however, no *post hoc* comparisons were statistically significant (**Figure 1D**, $p = 0.033$). Although not statistically significant, the largest changes in factorial aerobic scope occurred between 20 and 28°C ($p = 0.063$, $Q_{10} = 1.45$) and 28 and 32°C ($p = 0.084$, $Q_{10} = 0.47$), both of which resulted in a 26% decrease compared to 28°C.

## $U_{crit}$, $U_{opt}$, and Cost of Transport

Critical swimming speed ($U_{crit}$), a measurement intended to represent maximum sustained swimming ability (Tierney, 2011), followed a similar pattern to aerobic scope with the highest $U_{crit}$ occurring in fish acclimated to 28°C and a reduced $U_{crit}$ occurring at both higher and lower temperatures (**Figure 2A**, $p = 0.003$). $U_{crit}$ in animals acclimated to 20 and 32°C was significantly reduced compared to the 28°C-acclimated fish (28 and 23% decline, $p = 0.002$, $p = 0.022$, respectively). Optimal swimming speed ($U_{opt}$) was highest at 28°C but no statistically significant difference across temperatures was noted (**Figure 2B**; $p = 0.078$). $COT_{min}$ was significantly higher at 32°C compared to other temperatures (**Figure 3A**, $p < 0.001$). The $COT_{UCRIT}$ was significantly reduced at 20°C ($p < 0.001$), but no significant differences were noted between 24, 28, and 32°C (**Figure 3B**).

# DISCUSSION

The present study sought to investigate how acclimation to four different temperatures mahi experience in their natural environment (20, 24, 28, and 32°C) would affect their swimming performance and metabolic rate. Our results demonstrated that mahi performance was highest around 28°C, where we observed the largest aerobic scope and $U_{crit}$ (**Figures 1,2**). Results from this study generally agree with previous reported mahi distributions (see Introduction). Although fish in the present study were tested at an earlier age than wild-tagged adults, our results also tended to align with thermal ranges reported from PSAT data obtained from adult wild mahi (Schlenker et al., 2021). This, combined with the large habitat usage overlap across life stages of mahi (most time spent in surface waters), suggests that physiological performance observed in the present study may provide insight into thermal habitat utilization patterns of larger adults that are logistically difficult to test in laboratory settings.

Mortality noted in 32°C-acclimated fish suggests that the aerobic scope and $U_{crit}$ measurements obtained from the fish that completed the test are likely underestimating thermal impacts on performance at 32°C since we inadvertently selected for less temperature-sensitive fish. Notably, there was no mortality in any of the other acclimation temperatures. This trend has also been observed in earlier juvenile life stage mahi, where both $U_{crit}$ and metabolic rate were unaffected by acute 24 h exposure to 30°C, but only 63% of the exposed animals survived the exposure, and another ∼21% perished during introduction to swim tunnels or early in the swimming test (Mager et al., 2018). In another study on early juvenile life stage mahi, an acute heating challenge (0.83–1.09°C/h) resulted in frequent bottom-sitting and some mortality at 31.7–33°C (Szyper and Lutnesky, 1991). In the present study, 13 and 16% of mahi were disqualified from testing due to violation of the pre-established stipulations detailed in the methods section at 20 and 24°C, respectively, while no animals were disqualified at 28°C (**Supplementary Figure 1**). It is on this background that we discuss results from the current study.

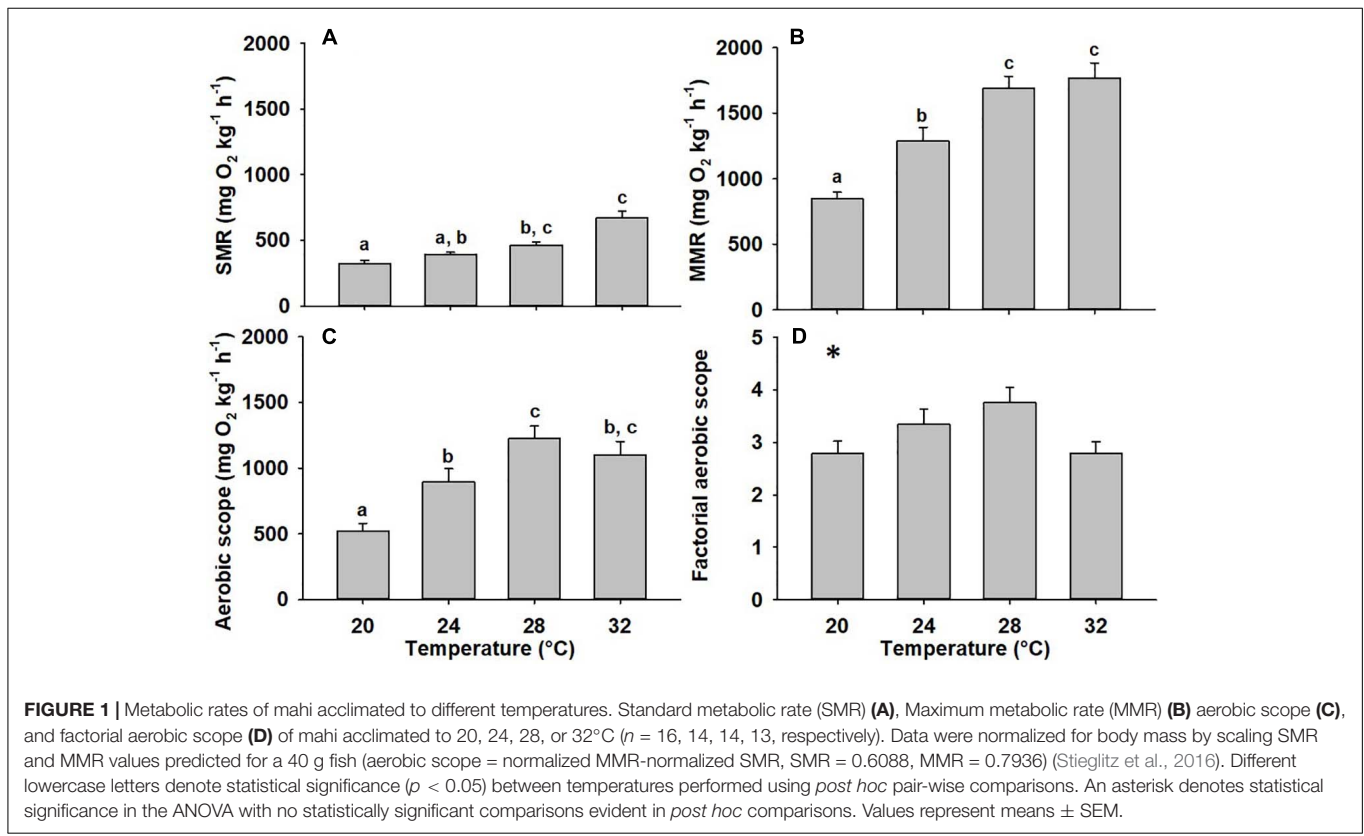

FIGURE 1 | Metabolic rates of mahi acclimated to different temperatures. Standard metabolic rate (SMR) (A), Maximum metabolic rate (MMR) (B) aerobic scope (C), and factorial aerobic scope (D) of mahi acclimated to 20, 24, 28, or 32°C ($n$ = 16, 14, 14, 13, respectively). Data were normalized for body mass by scaling SMR and MMR values predicted for a 40 g fish (aerobic scope = normalized MMR-normalized SMR, SMR = 0.6088, MMR = 0.7936) (Stieglitz et al., 2016). Different lowercase letters denote statistical significance ($p < 0.05$) between temperatures performed using *post hoc* pair-wise comparisons. An asterisk denotes statistical significance in the ANOVA with no statistically significant comparisons evident in *post hoc* comparisons. Values represent means ± SEM.

As expected, standard metabolic rate increased with temperature (**Figure 1A**) and the largest incremental increase occurred between 28 and 32°C ($Q_{10}$ = 2.49). Although not statistically significant, this 44% increase in metabolic rate suggested that baseline metabolic processes are costly in 32°C-acclimated animals. Maximum metabolic rate also increased with temperature but plateaued at higher temperatures (**Figure 1B**). Due to the changes in standard metabolic rate and maximum metabolic rate, absolute aerobic scope was significantly impacted by temperature (**Figure 1C**). Aerobic scope was highest at 28°C and declined in either direction from this temperature. Animals acclimated to 20°C had the lowest aerobic scope that appeared to result from an inability to increase maximum metabolic rate. Driven largely by increased standard metabolic rate, mahi acclimated to 32°C experienced a non-significant 10% decline in aerobic scope compared to 28°C, but this result may be conservative due to treatment-induced mortality and disqualifications (see above). Factorial aerobic scope showed a similar pattern to absolute aerobic scope, but with a larger (26%) but not statistically significant decrease ($p$ = 0.084) when comparing 28 and 32°C acclimated mahi ($Q_{10}$ = 0.47). The reduction in aerobic scope in 20°C and trend for reduction in 32°C acclimated animals could suggest that oxygen supply and demand may be mismatched at these temperatures compared to 28°C, however, the specific mechanisms responsible for these declines in performance was not examined in the present study. Changes in hematocrit, cardiac output, mitochondrial density, mitochondrial function, contractile protein function, electrophysiological properties of cardiac and muscle cells, adrenergic responses, cardiac myoglobin levels, relative ventral mass, gill perfusion, and/or gill structural changes could all be contributing to these differences (Egginton and Sidell, 1989; Keen and Farrell, 1994; Gallaugher et al., 1995; Watabe, 2002; Shiels et al., 2003, 2004; Sollid and Nilsson, 2006; Pörtner and Knust, 2007; Eliason et al., 2011; Anttila et al., 2013; Farrell, 2016), and are potential avenues to pursue in future studies.

In one study that has examined cardiovascular parameters as a function of temperature, mahi embryos raised at 30°C had significantly higher heart rates than embryos raised at 26°C (Perrichon et al., 2017). In an isolated preparation, the contractile force of ventricular heart muscle from adult mahi was measured during acute cooling from 26 to 7°C (Galli et al., 2009). Interestingly, contractile force increased and peaked at 18.2°C and then declined as temperatures continued to cool. Addition of adrenaline increased contractile force as expected, but generally showed a similar pattern of decline under ∼18°C. These results suggest that cardiac adjustments are limited at cooler temperatures close to the 20°C used in the present study. Pharmacological blockade of the sarcoplasmic reticulum (SR) did not change contractile force between ∼21–26°C, but was effective at temperatures below 20°C, suggesting that mahi have the ability to recruit additional intracellular calcium needed for contraction at low temperatures, but are generally less reliant on the SR for contraction between ∼21–26°C (Galli et al., 2009). Temperatures below 20°C where SR recruitment would occur, would likely only be transient based on behavior of satellite-tagged wild mahi (Merten et al., 2014b; Schlenker et al., 2021). Another recent study examining mitochondrial function in adult mahi reared close to 26°C and assayed at 20, 26, and 30°C, found that heart mitochondria were more sensitive when compared to those from the red muscle, and showed reduced phosphorylation efficiency as temperatures increased (Lau et al., 2020). Examining contractile force, adrenergic responses, and/or SR-dependence at warmer temperatures, as well as heart mitochondrial function in temperature-acclimated animals, would be useful future avenues of research, to determine if declines in cardiac function also underlie reduced swimming performance at 32°C observed in the present study.

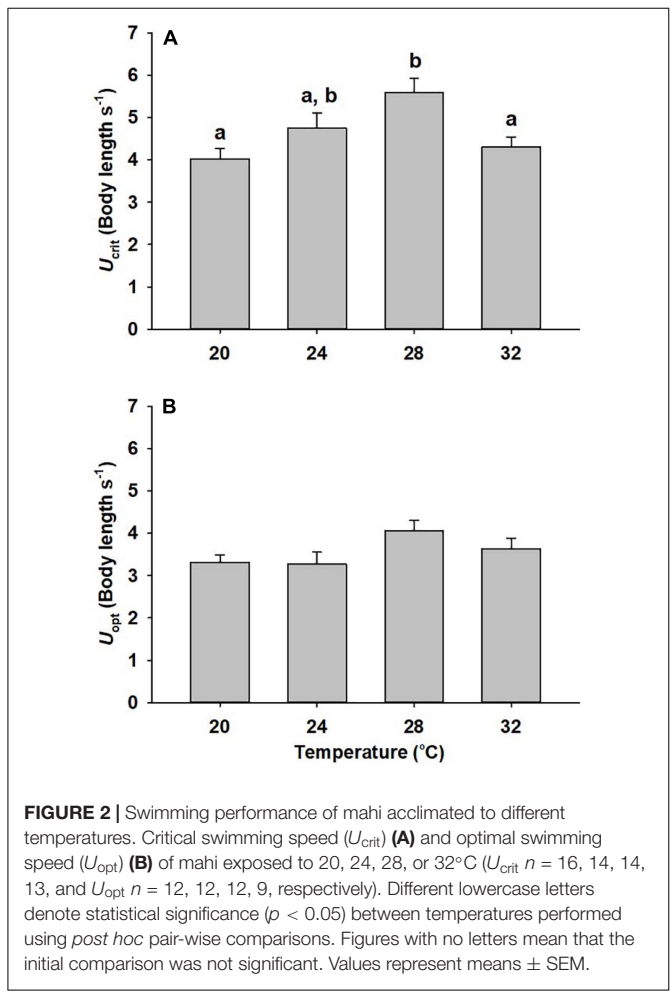

FIGURE 2 | Swimming performance of mahi acclimated to different temperatures. Critical swimming speed ($U_{crit}$) **(A)** and optimal swimming speed ($U_{opt}$) **(B)** of mahi exposed to 20, 24, 28, or 32°C ($U_{crit}$ $n$ = 16, 14, 14, 13, and $U_{opt}$ $n$ = 12, 12, 12, 9, respectively). Different lowercase letters denote statistical significance ($p < 0.05$) between temperatures performed using *post hoc* pair-wise comparisons. Figures with no letters mean that the initial comparison was not significant. Values represent means ± SEM.

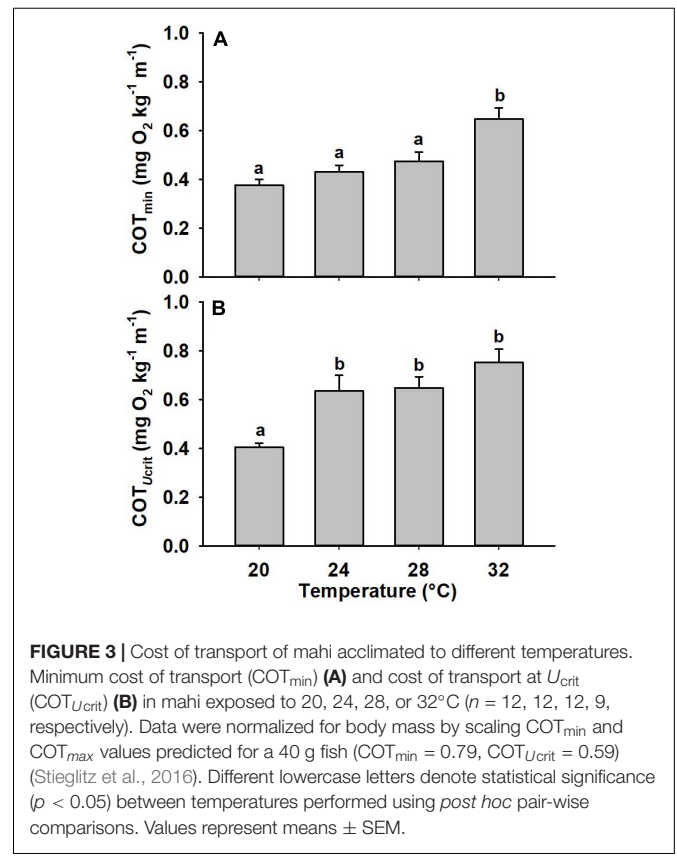

FIGURE 3 | Cost of transport of mahi acclimated to different temperatures. Minimum cost of transport ($COT_{min}$) **(A)** and cost of transport at $U_{crit}$ ($COT_{Ucrit}$) **(B)** in mahi exposed to 20, 24, 28, or 32°C ($n$ = 12, 12, 12, 9, respectively). Data were normalized for body mass by scaling $COT_{min}$ and $COT_{max}$ values predicted for a 40 g fish ($COT_{min}$ = 0.79, $COT_{Ucrit}$ = 0.59) (Stieglitz et al., 2016). Different lowercase letters denote statistical significance ($p < 0.05$) between temperatures performed using *post hoc* pair-wise comparisons. Values represent means ± SEM.

As expected, critical swimming speed ($U_{crit}$) followed a similar pattern to aerobic scope with the highest $U_{crit}$ occurring in fish acclimated to 28°C and a significantly reduced $U_{crit}$ occurring at both the highest and the lowest temperatures (**Figure 2A**). Although there were differences in $U_{crit}$, optimal swimming speed remained similar across temperatures, suggesting that mahi have a relatively fixed optimal swimming speed regardless of temperature, but for 20 and 32°C-acclimated animals, this optimum is closer to $U_{crit}$. $COT_{min}$ and the cost of transport at $U_{crit}$ ($COT_{UCRIT}$) generally showed an increase with temperature reflecting changes in standard metabolic rate.

Our results align with previous field observations, including fisheries data that indicates higher catch rates between 24 and 29°C and from popup satellite archival data that indicate mahi spend most of their time between ∼25–29°C (Flores et al., 2008; Hammond, 2008; Martínez-Rincón et al., 2009; Farrell et al., 2014; Merten et al., 2014b; Schlenker et al., 2021). Around the world, mahi are generally found above 20°C, which may explain why we noted reduced aerobic scope, $U_{crit}$, and maximum metabolic rate at lower temperatures (Palko et al., 1982). Assuming that thermal performance is similar across life stages, the drastic declines in performance noted at 20°C in the present study suggest that mahi may have a limited capacity to function during long periods of high activity, potentially making it difficult for them to catch prey and avoid predation. This is further supported when noting that the highest calculated $Q_{10}$ of all metrics and temperatures in this study occurred at the transition from 20 to 24°C for absolute aerobic scope ($Q_{10}$ = 3.83), driven mostly by maximum metabolic rate ($Q_{10}$ = 2.86). Fisheries data indicate that the coolest months (∼20–21°C) result in the lowest catch rates in the Gulf of California, Mexico (Flores et al., 2008). In a comparison between Pacific bluefin tuna (*Thunnus orientalis*) and mahi residing in the same area in the East China Sea, mahi remained above the thermocline (generally above 20°C), while tuna frequently crossed the thermocline into cooler waters (16°C). As the thermocline became deeper over time, mahi extended their vertical range. This, in combination with southern migration of mahi with the 20°C isotherm, suggested that the vertical movements were constrained due to low temperatures, although authors could not rule out that these observations were due to prey availability or predator avoidance (Furukawa et al., 2014). In a study from the western Atlantic, one tagged mahi was found at 16°C for short periods of time, but most individuals stayed above 19°C, in agreement with studies from other regions (Merten et al., 2014b).

The increase in mortality, reduced $U_{crit}$, increased standard metabolic rate, and a trend toward a lower absolute and factorial aerobic scope at 32°C suggests that this temperature may limit oxygen-demanding activity in mahi. The highest aerobic scope and $U_{crit}$ in the present study from late juvenile stage fish at 28°C is interesting when compared to PSAT data from wild-tagged adult mahi in a recent study of 19 fish in the western Atlantic and the Gulf of Mexico (Schlenker et al., 2021). Despite having access to a range of temperatures, mahi in this study spent most of their time at 27.5°C. Although temperatures between 28 and 32°C were not tested in the present study, the comparison of optimal performance in the lab and median PSAT temperatures could suggest that mahi are utilizing their habitat in a manner that maximizes their aerobic capacity near 28°C, if one assumes that high levels of habitat overlap across life stages means thermal performance is similar. Our results tend to agree with earlier studies demonstrating the utility of aerobic scope to predict performance and occurrence in the wild (Pörtner and Knust, 2007; Farrell et al., 2008; Eliason et al., 2011) although we note that there may be exceptions to this relationship between aerobic scope and performance (Gräns et al., 2014; Norin et al., 2014; Jutfelt et al., 2018; Slesinger et al., 2019; Seibel and Deutsch, 2020). Examination of hypoxia tolerance ($P_{crit}$) could provide further insight into the mechanism and provide an independent test of oxygen supply capacity theory (Deutsch et al., 2020; Seibel and Deutsch, 2020). Of further interest is that although average sea surface temperatures in the Gulf of Mexico were warmer than in the western Atlantic, wild-tagged mahi maintained similar median temperatures across both regions. One key difference between fish tagged in the two regions is that Gulf of Mexico-tagged mahi resided at a deeper average depth, where waters were cooler (Schlenker et al., 2021). This could demonstrate that mahi are already altering habitat utilization patterns to live in areas where average sea surface temperatures are higher.

Above, we compare observations from mahi tested at an earlier age in the laboratory to PSAT tagged adults in the wild. A recent metanalysis has demonstrated that embryonic and spawning adults may be more thermally sensitive compared to juveniles and pre-spawning adults (Dahlke et al., 2020). However, mahi were not included in this analysis and all mahi life stages are highly surface oriented and found in the same geographical areas, suggesting overlapping thermal habitats, and thus, potentially similar thermal tolerances. Although less is known about the life stage used in the current study with respect to habitat utilization in the wild, fish at the age and size used in the present study (**Table 1**) have a fully forked tail, exhibit adult coloration pattern and schooling behavior, and are found in habitats where adults are regularly present (Oxenford, 1999; Perrichon et al., 2019; Schlenker et al., 2021). Nonetheless, insight into thermal responses and measurements of thermal plasticity at more life stages would be useful to make more targeted predictions regarding climate change impacts (Schulte et al., 2011; Dahlke et al., 2020; Havird et al., 2020). In addition, food was provided *ad libitum* in this study, which may not necessarily reflect food availability in the wild at different temperatures.

Although mahi are highly migratory and readily travel long distances, increasing temperatures expected due to climate change could alter habitat utilization patterns both directly, by constraining aerobic scope, or indirectly, by altering the distribution of their prey. For example, mahi are found throughout the year in the Gulf of Mexico, where mean summer sea surface temperatures currently are 29.5°C. By year 2100, recent modeling has indicated this average summer sea surface temperature is expected to climb to more than 32°C (Laurent et al., 2018). Given that mahi currently only spend a small portion of their time in 32°C water despite periodically experiencing it in their environment, these projections for Gulf of Mexico sea surface temperatures suggest that mahi may not thrive during the summer months in a future and warmer Gulf of Mexico. They could alter their distribution, both geographically and vertically in the water column. Shifts in latitudinal distributions due to increased temperatures have already been noted for other fish species (Perry et al., 2005; Nicolas et al., 2011; Fossheim et al., 2015) and have been predicted specifically for mahi along the North American west coast using Ecological Niche Modeling (Salvadeo et al., 2020).

Warmer waters will have reduced oxygen concentrations. For example in the Gulf of Mexico, higher temperatures are predicted to result in 3.4–9.4% lower oxygen concentrations in the summer, depending on depth (Laurent et al., 2018). Given the temperature-induced declines in performance noted in the present study and previous observations of hypoxia-induced reductions in $U_{crit}$ in mahi (Mager et al., 2018), synergistic interactions between warming temperatures and declining oxygen concentrations may impact migratory fish, such as mahi, by constraining geographical range and/or forcing them to alter habitat utilization patterns.

In conclusion, although performance of mahi has been extensively studied following the *Deepwater Horizon* oil spill in the Gulf of Mexico (Mager et al., 2014; Nelson et al., 2016; Stieglitz et al., 2016; Pasparakis et al., 2019), little was known about how basic environmental factors such as temperature impact swimming performance and aerobic capacity in this species. The present study reveals important information regarding temperature and performance of mahi in a laboratory setting, and may provide insight into habitat utilization patterns observed from fisheries catch data and from satellite-tagged wild mahi at later life stages. This study also suggests that using lab-based performance data to predict wild fish patterns can be an asset when considering both current and future habitat utilization patterns. Future studies examining the physiological mechanisms that dictate cold and warm tolerance limits in mahi would add to understanding of how this species may respond to climate change (Chen et al., 2015).

## DATA AVAILABILITY STATEMENT

The datasets presented in this study can be found in online repositories. The names of the repository/repositories and accession number(s) can be found below: https://data.gulfresearchinitiative.org/data/R6.x804.000:0008; Gulf of Mexico Research Initiative Information and Data Cooperative (GRIIDC) (Heuer et al., 2020; doi: 10.7266/n7-xd7p-yj29).

## ETHICS STATEMENT

The animal study was reviewed and approved by University of Miami's Institutional Animal Care and Use Committee.

## AUTHOR CONTRIBUTIONS

RH, JS, and MG conceptualized the work. RH, JS, IE, and MG contributed to the development and/or design of methodology. RH performed the experiments and analyzed the data. CP assisted with data acquisition. JS, IE, DB, and MG provided resources. MG supervised the project and procured funding for the project. RH wrote the manuscript and all co-authors reviewed and edited the manuscript. All authors contributed to the article and approved the submitted version.

## FUNDING

This research was supported by a grant from the Gulf of Mexico Research Initiative [Grant No. SA-1520 to the RECOVER consortium (Relationship of effects of cardiac outcomes in fish for validation of ecological risk)].

## ACKNOWLEDGMENTS

We would like to thank Lela S. Schlenker, Rebecca L. Zlatkin, Emma Esch, and Gabrielle Menard for assistance with water quality monitoring and fish care during experiments. We would also like to thank Derek Manzello, Leah Chomiak, Graham Kolodziej, Paul Jones, John Morris, and Nathan Formel for providing use of and assistance with experimental tanks (CIMAS, NOAA and AOML, as well as the ACCRETE group). We would also like to acknowledge NOAA's Omics Initiative for funding the analytical equipment necessary for tank conditions. MG is a Maytag Professor of Ichthyology. Finally, we would like to thank the University of Miami Experimental Hatchery (UMEH) for maintaining the mahi.

## SUPPLEMENTARY MATERIAL

The Supplementary Material for this article can be found online at: https://www.frontiersin.org/articles/10.3389/fmars.2021.654276/full#supplementary-material

## Supplementary Material

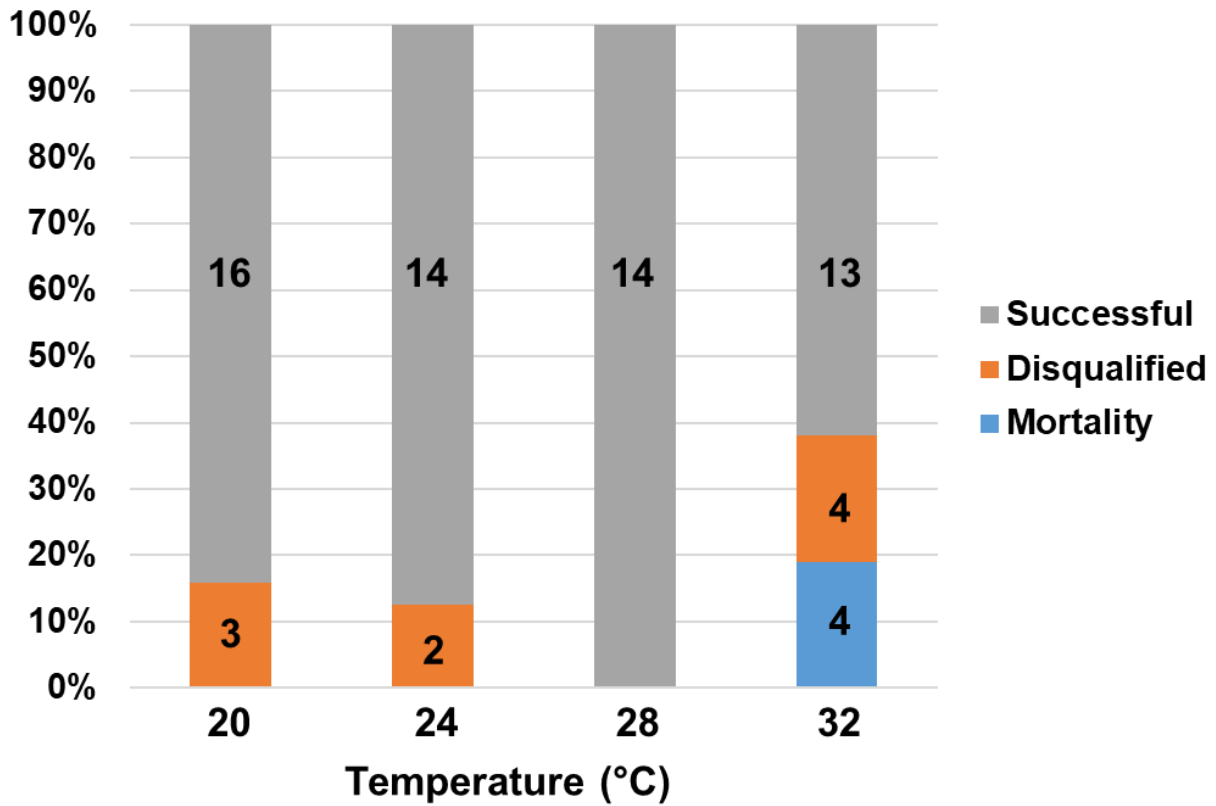

**Supplementary Figure S1:** Comparisons of successful swim tunnel respirometry attempts (grey), disqualified (orange), and mortality (blue) as a function of total number of attempts scaled to 100% for each temperature. Numbers within each respective bar, represent the total n-number in each respective scenario. Criteria for disqualification (orange) included animals with poor $r^2$ in plots illustrated in Supplementary Figure S4 (<0.70) and/or animals that refused to swim during the test.

*Supplementary Material*

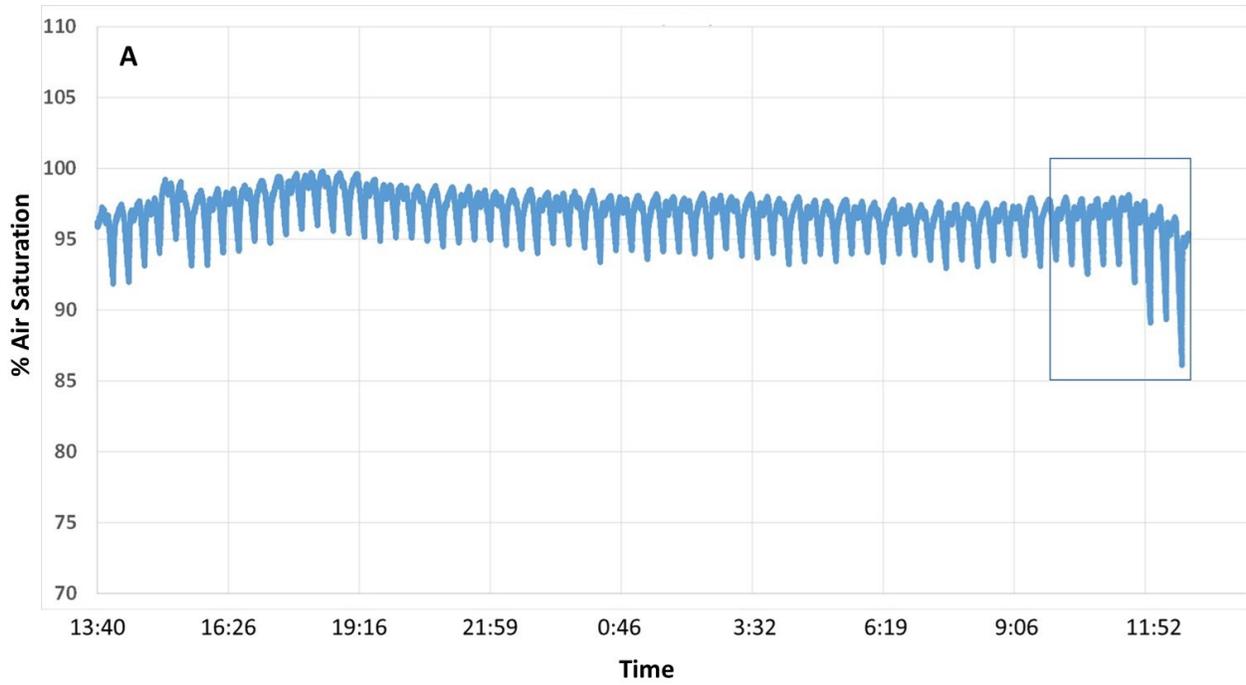

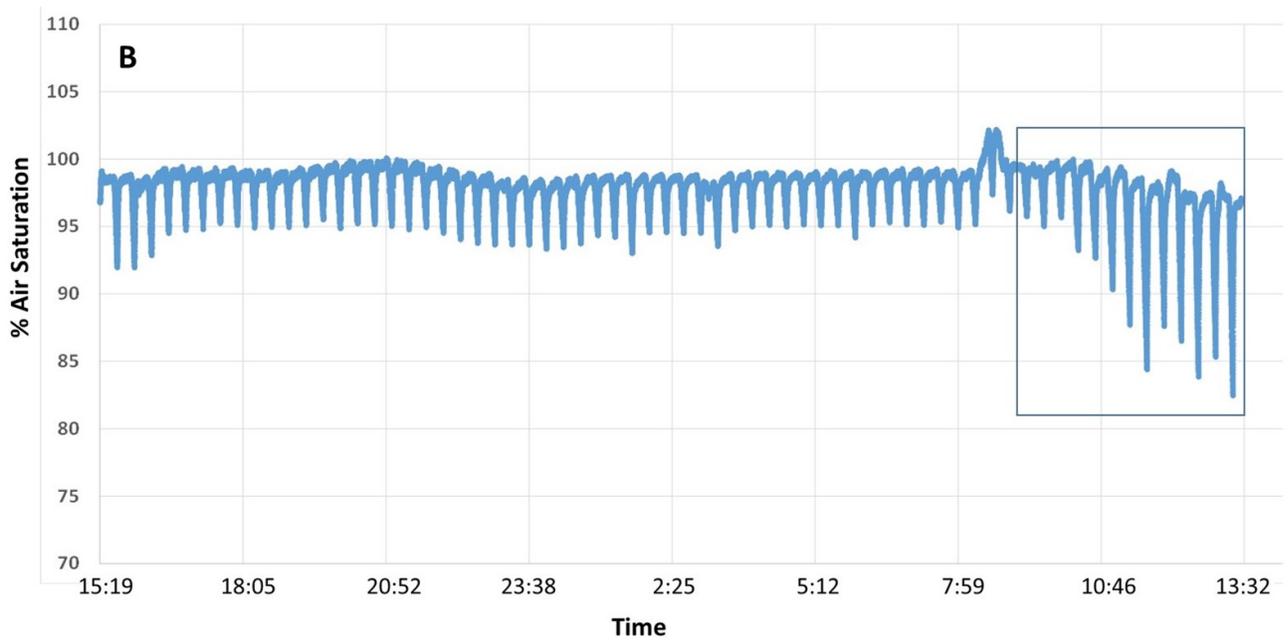

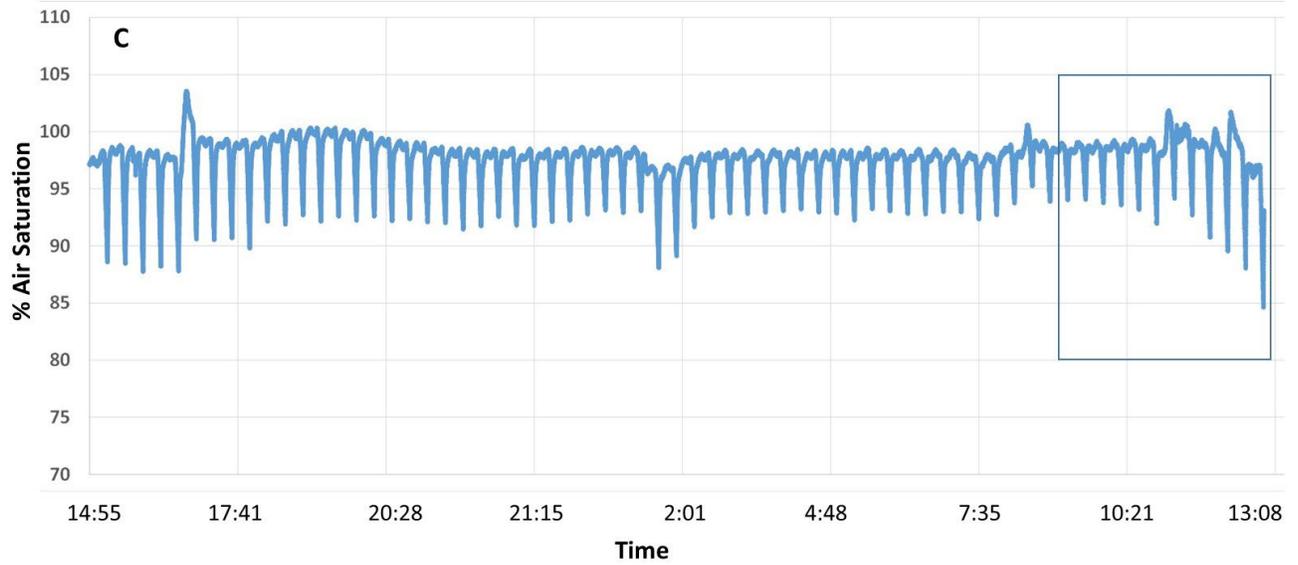

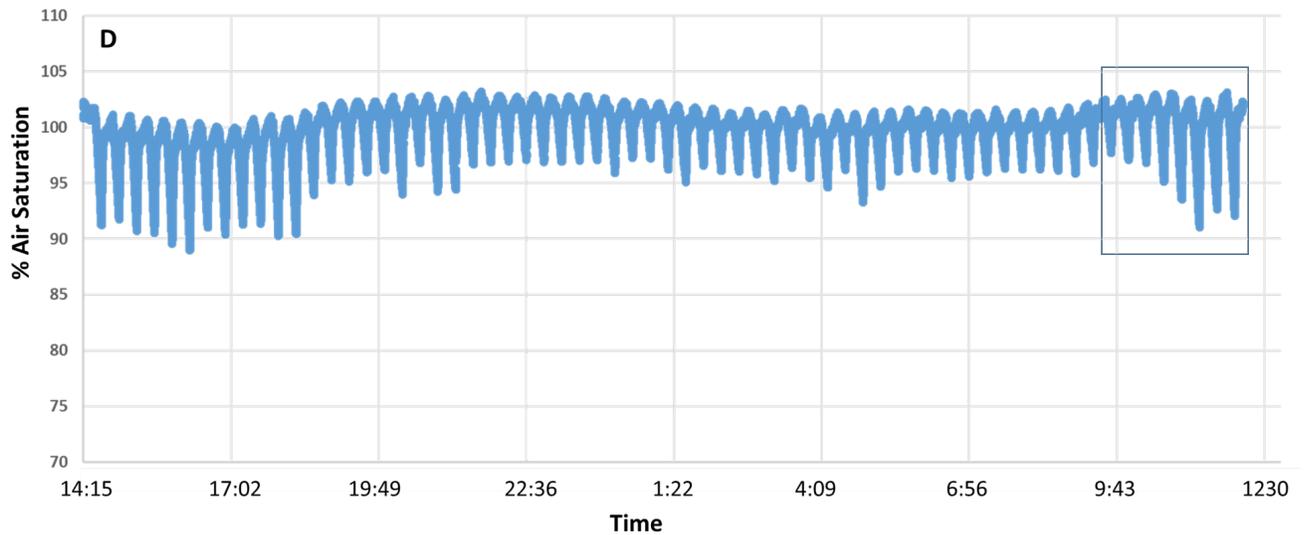

**Supplementary Figure S2**: A representative trace showing oxygen consumption of mahi over time acclimated to (A) 20°C, (B) 24°C, (C) 28°C, (D) 32°C. The blue box represents the time period covering the critical swim test ($U_{crit}$).

# *Supplementary Material*

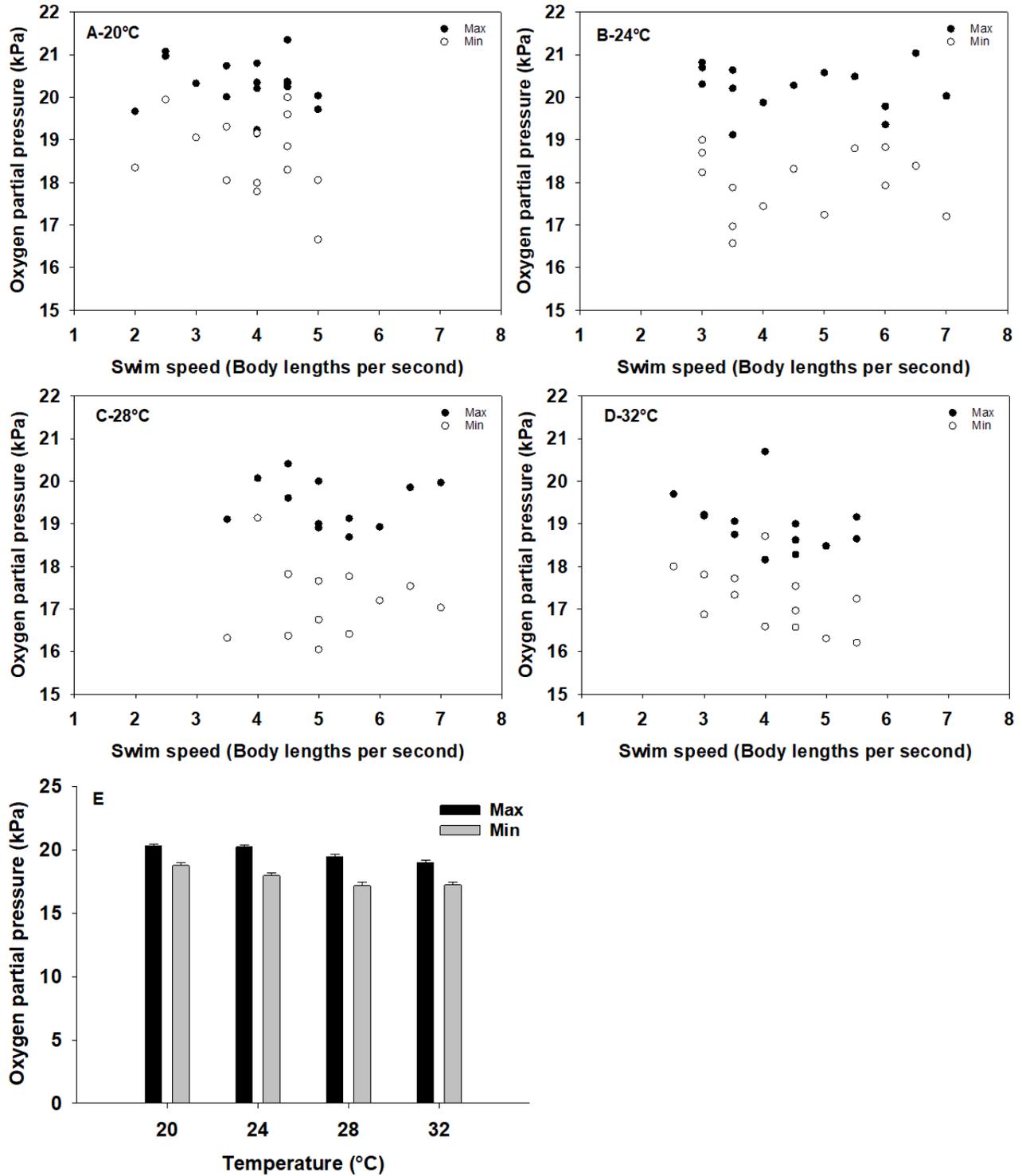

**Supplementary Figure S3**: **Maximum, minimum, and average oxygen partial pressure (kPa) as a function of temperature and swim speed.** Data are from interval prior to failure interval during $U_{crit}$ (see text for calculation of $U_{crit}$) Panels (**A-D**) show maximum and minimum oxygen partial pressure in individuals acclimated to 20, 24, 28, and 32°C, respectively. (**E**) Means ± standard error of maximum and minimums presented in panels A-D.

## *Supplementary Material*

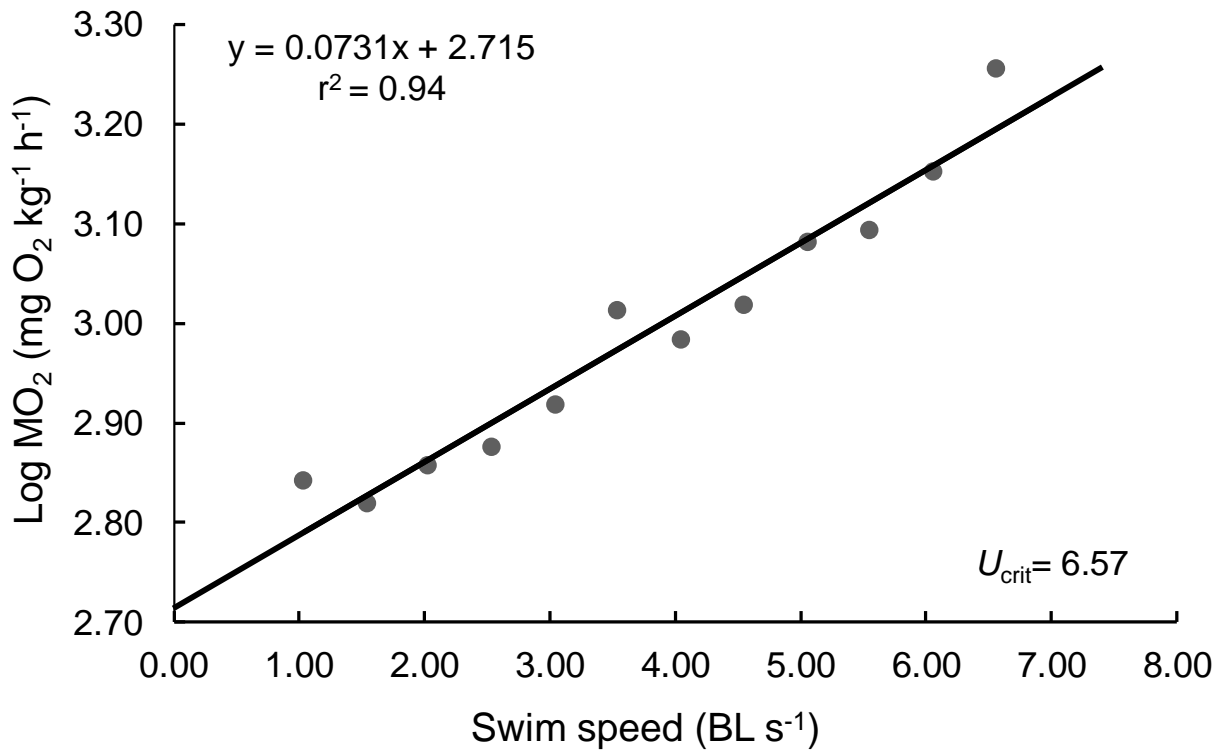

**Supplementary Figure S4**: A sample least squares regression of metabolic rate versus swimming speed (*U*) from an individual Mahi-mahi following a swim respirometry trial. Standard metabolic rate (SMR) is the y-intercept and maximum metabolic rate (MMR) is the metabolic rate extrapolated to $U_{crit}$. Aerobic scope was MMR-SMR and only regressions with an $r^2 \geq 0.7$ were used in analyses. Data were normalized for body mass by scaling SMR and MMR values predicted for a 40g fish using scaling coefficients previously recorded for this species (aerobic scope=normalized MMR - normalized SMR, SMR=0.6088, MMR=0.7936) (Stieglitz et al., 2016).

**Reference:**

Stieglitz, J.D., Mager, E.M., Hoenig, R.H., Benetti, D.D., and Grosell, M. (2016). Impacts of Deepwater Horizon crude oil exposure on adult mahi-mahi (Coryphaena hippurus) swim performance. *Environmental toxicology and chemistry* 35(10)**,** 2613-2622.

## Supplementary Material

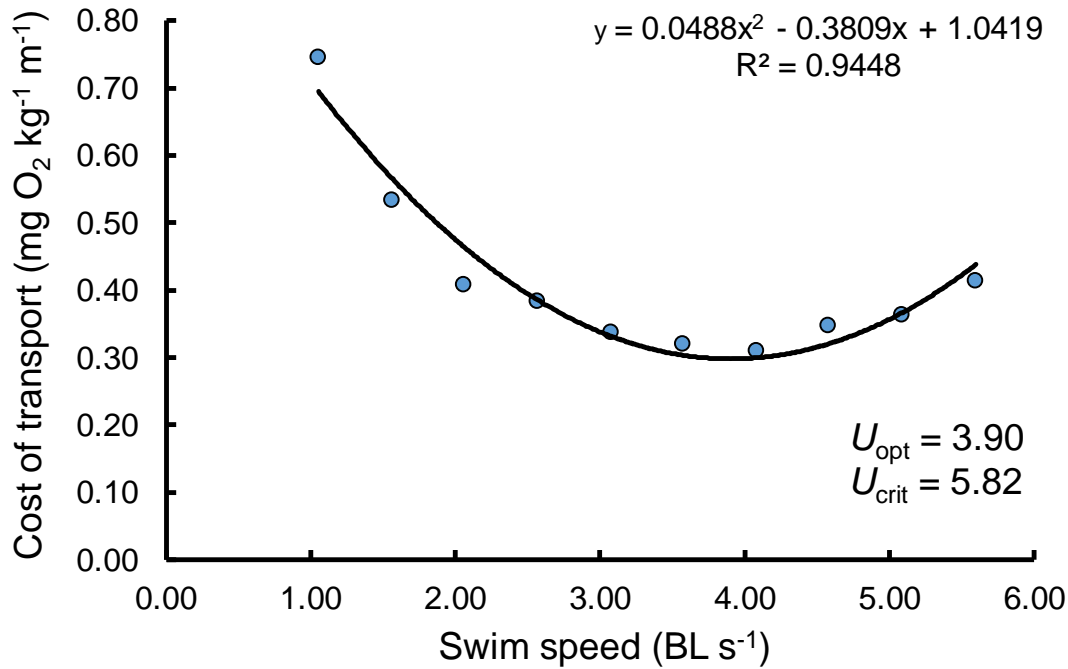

**Supplementary Figure S5**: A sample cost of transport plot showing cost of transport (COT) plotted against swimming speed ($U$) from an individual Mahi-mahi following a swim respirometry trial. This plot provided minimum cost of transport ($COT_{min}$), the cost of transport at $U_{crit}$ ($COT_{Ucrit}$), in addition to the optimal swimming speed ($U_{opt}$), which is the speed at which swimming required the minimum cost of transport and was determined by fitting the first derivative of the polynomial to zero (Palstra et al., 2008; Stieglitz et al., 2016) Only individuals with an $r^2$ greater than or equal to 0.7 were used in analyses. Mass was normalized to a 40g fish using previously determined scaling coefficients in this species for minimum cost of transport and cost of transport at $U_{crit}$ ($COT_{min} = 0.79$, $COT_{UCrit} = 0.59$) (Stieglitz et al., 2016).

# Supplementary Material

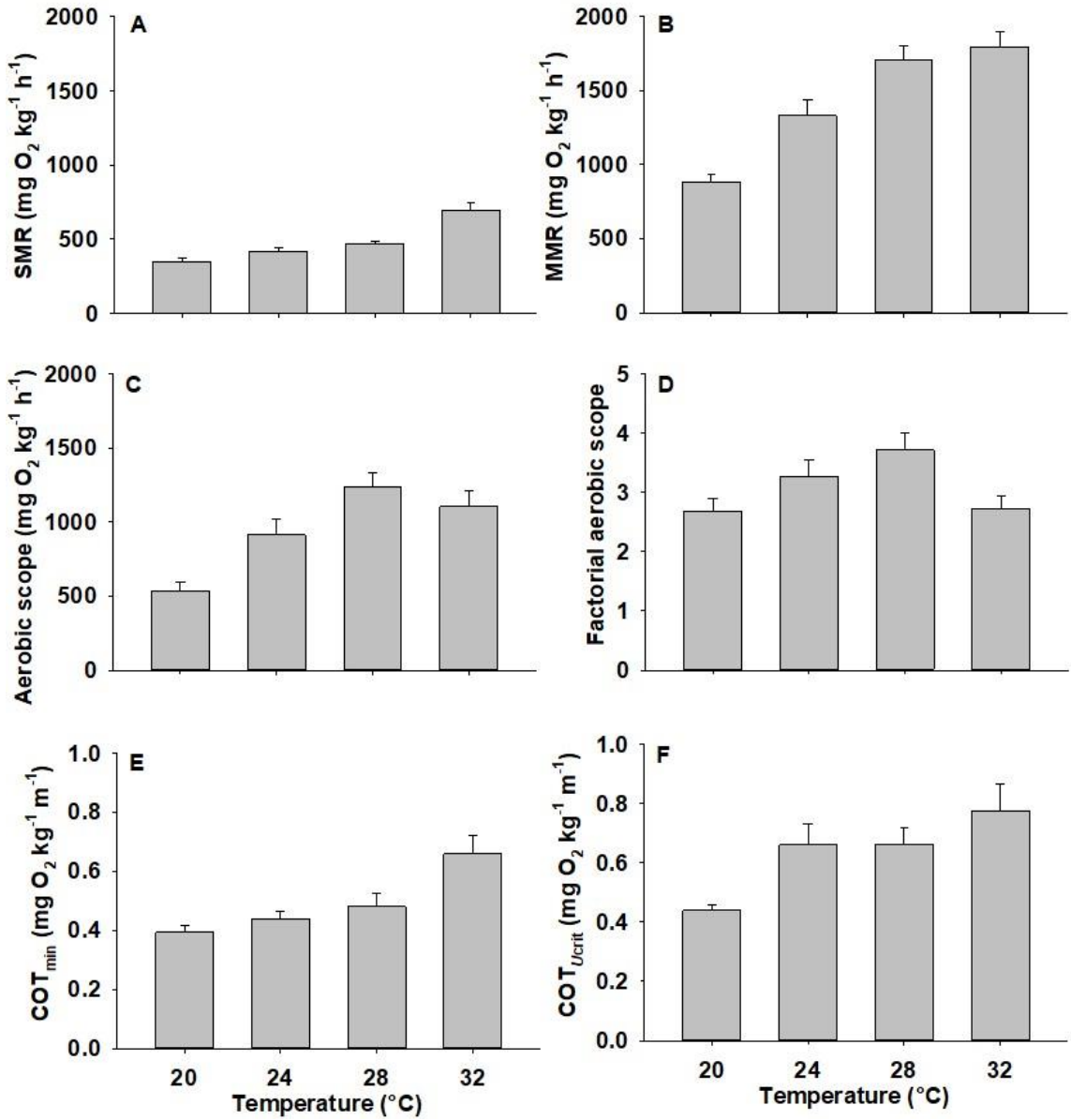

**Supplementary Figure S6**: **Metabolic rates and cost of transport of mahi acclimated to different temperatures with no mass scaling coefficients.** Standard metabolic rate (SMR) (**A**), Maximum metabolic rate (MMR) (**B**), aerobic scope (**C**), factorial aerobic scope (**D**), minimum cost of transport (**E**), and cost of transport at $U_{crit}$ (**F**) of mahi acclimated to 20, 24, 28, or 32ºC.

## *Supplementary Material*

**Supplementary Table S1**: Water chemistry parameters for each temperature treatment (20, 24, 28, or 32ºC). Averages reflect means of multiple tank replicates (20ºC: n=3, 24ºC: n=4, 28ºC: n=5, 32ºC: n=6). Values are presented as means ± s.e.m.

| Temperature Treatment | $pH_{NBS}$ | Temperature (ºC) | Salinity | Dissolved oxygen (mg/L) | Ammonia (µM) |
|---|---|---|---|---|---|
| 20ºC | 8.05 ± 0.02 | 20.2 ± 0.02 | 35.3 ± 0.7 | 7.33 ± 0.10 | 6.7 ± 1.2 |
| 24ºC | 8.04 ± 0.01 | 24.1 ± 0.03 | 35.3 ± 0.5 | 6.90 ± 0.02 | 6.9 ± 1.2 |
| 28ºC | 8.02 ± 0.01 | 28.0 ± 0.03 | 35.1 ± 0.5 | 6.58 ± 0.10 | 8.9 ± 2.0 |
| 32ºC | 8.03 ± 0.01 | 32.0 ± 0.04 | 35.9 ± 0.5 | 6.20 ± 0.09 | 8.6 ± 1.5 |

# Supplementary Material

**Supplementary Table S2**: Temperature coefficients ($Q_{10}$) at increasing temperature (20-24, 24-28, 28-32°C) for standard metabolic rate, maximum metabolic rate, aerobic scope, critical swimming speed ($U_{crit}$), optimal swimming speed, minimum cost of transport ($COT_{min}$), and cost of transport at $U_{crit}$ ($COT_{UCRIT}$). $Q_{10}$ was calculated using the formula $Q_{10}=(k_2-k_1)^{10/(t2-t1)}$, where $k_1$ and $k_2$ are the rate constants at $t_1$ and $t_2$, respectively (Kieffer et al., 1998). See primary text for units associated with each endpoint.

| Measured endpoint | Temperature (°C) | $Q_{10}$ |
|---|---|---|
| Standard metabolic rate | 20-24°C | 1.63 |
|  | 24-28°C | 1.51 |
|  | 28-32°C | 2.49 |
| Maximum metabolic rate | 20-24°C | 2.86 |
|  | 24-28°C | 1.97 |
|  | 28-32°C | 1.12 |
| Aerobic scope | 20-24°C | 3.83 |
|  | 24-28°C | 2.19 |
|  | 28-32°C | 0.76 |
| $U_{crit}$ | 20-24°C | 1.51 |
|  | 24-28°C | 1.50 |
|  | 28-32°C | 0.52 |
| $U_{opt}$ | 20-24°C | 0.98 |
|  | 24-28°C | 1.69 |
|  | 28-32°C | 0.77 |
| $COT_{min}$ | 20-24°C | 1.39 |
|  | 24-28°C | 1.28 |
|  | 28-32°C | 2.17 |
| $COT_{UCRIT}$ | 20-24°C | 3.08 |
|  | 24-28°C | 1.06 |
|  | 28-32°C | 1.45 |
| Factorial aerobic scope | 20-24°C | 1.57 |
|  | 24-28°C | 1.34 |
|  | 28-32°C | 0.47 |
|  | 20-28°C | 1.45 |

## *Supplementary Material*

**Supplementary Table S3**: Results of statistical analyses on mahi metabolic rates and swim parameters acclimated to either 20, 24, 28, or 32ºC. Significance for all statistical tests was determined at p<0.05, and means are presented ± SEM. See main text for sample sizes.

| Measured variable | Statistical test | P-value | Post-hoc test | Post-hoc test P-values | Median (Range) |
|---|---|---|---|---|---|
| Standard metabolic rate | Kruskall-Wallace ANOVA on ranks | P<0.001 | Dunn's | 32 v 20 = <0.001<br>32 v 24 = 0.002<br>32 v 28 = 0.219<br>28 v 20 = 0.007<br>28 v 24 = 0.815<br>24 v 20 = 0.527 | 20ºC: 304.5 (218.7-507.6)<br>24ºC: 387.1 (315.8-514.0)<br>28ºC: 452.5 (344.6-645.6)<br>32ºC: 644.6 (354.4-1016.8) |
| Maximal metabolic rate | Welch's ANOVA | P<0.001 | Games-Howell | 20 v 24 = 0.004<br>20 v 28 = <0.001<br>20  32 = <0.001<br>24 v 28 = 0.031<br>24 v 32 = 0.020<br>28 v 32 = 0.949 | 20ºC: 829.5 (631.0-1260.3)<br>24ºC: 1228.7 (854.6-2235.2)<br>28ºC: 1685.8 (1202.1-2280.8)<br>32ºC: 1799.5 (1180.7-2347.2) |
| Aerobic scope | One-way ANOVA | P<0.001 | Holm-Sidak | 28 v 20 = <0.001<br>32 v 20 = <0.001<br>24 v 20 = 0.014<br>28 v 24 = 0.032<br>32 v 24 = 0.217<br>28 v 32 = 0.328 | 20ºC: 503.4 (179.0-973.7)<br>24ºC: 846.0 (444.3-1877.4)<br>28ºC: 1122.4 (781.8-1936.2)<br>32ºC: 980.7 (565.1-1775.4) |
| Critical swimming speed ($U_{crit}$) | One-way ANOVA | P=0.003 | Holm-Sidak | 28 v 20 = 0.002<br>28 v 32 = 0.022<br>28 v 24 = 0.194<br>24 v 20 = 0.238<br>24 v 32 = 0.528<br>32 v 20 = 0.512 | 20ºC: 4.191 (1.991-5.445)<br>24ºC: 4.461 (3.222-6.982)<br>28ºC: 5.463 (3.622-8.359)<br>32ºC: 4.301 (3.050-5.709) |
| Optimal sustained swimming speed ($U_{opt}$) | One-way ANOVA | P=0.078 | N/A | N/A | 20ºC: 3.480 (2.440-4.510)<br>24ºC: 3.140 (2.170-4.710)<br>28ºC: 3.975 |
| | | | | | (2.560-5.340) |
| | | | | | 32ºC: 3.700 (2.470-4.630) |
| Minimum cost of transport (COT$_{min}$) | One-way ANOVA | P<0.001 | Holm-Sidak | 32 v 20 = <0.001<br>32 v 24 = <0.001<br>32 v 28 = 0.005<br>28 v 20 = 0.110<br>24 v 20 = 0.447<br>28 v 24 = 0.327 | 20ºC: 0.379 (0.292-0.541)<br>24ºC: 0.404 (0.315-0.616)<br>28ºC: 0.470 (0.315-0.792)<br>32ºC: 0.665 (0.426-0.891) |
| Cost of transport at $U_{crit}$ (COT$_{UCRIT}$) | Kruskall-Wallace ANOVA on ranks | P<0.001 | Dunn's | 32 v 20 = <0.001<br>32 v 24 = 0.792<br>32 v 28 = 1.000<br>28 v 20 = 0.002<br>28 v 24 = 1.000<br>24 v 20 = 0.009 | 20ºC: 0.410 (0.322-0.520)<br>24ºC: 0.551 (0.373-1.113)<br>28ºC: 0.617 (0.451-0.893)<br>32ºC: 0.734 (0.546-1.027) |
| Factorial aerobic scope | Kruskall-Wallace ANOVA on ranks | P=0.033 | Dunn's | 32 v 20 = 1.000<br>32 v 24 = 1.000<br>32 v 28 = 0.084<br>28 v 20 = 0.063<br>28 v 24 = 1.000<br>24 v 20 = 1.000 | 20ºC: 2.704 (1.390-4.398)<br>24ºC: 3.238 (2.079-6.246)<br>28ºC: 3.541 (2.572-6.618)<br>32ºC: 2.835 (1.816-4.609) |